\documentclass[11pt]{article}
\usepackage{multicol}
\usepackage[margin=0.75in]{geometry}
\usepackage{amsmath}       
\usepackage{amssymb}       
\usepackage{bm}            
\usepackage{graphicx}      
\usepackage[font=small,labelfont=bf]{caption}  
\usepackage{hyperref}      
\usepackage{cite}

\title{Two Complementary Approaches to Determine the Effective Index of Dielectric Ring Resonators}
\author{Ergun Simsek \\
Department of Computer Science and Electrical Engineering,\\ 
University of Maryland Baltimore County, Baltimore, MD, 20817, United States \\
simsek@umbc.edu}

\begin{document}
\maketitle
\abstract{
Open-source finite-element frameworks for modal analysis of dielectric ring resonators are presented based on two complementary approaches. The first computes resonant wavelengths for discrete azimuthal mode numbers and obtains the effective index at a target wavelength by interpolation; the second solves directly for the azimuthal mode number at a prescribed wavelength via a quadratic eigenvalue formulation. Both approaches are validated against a commercial electromagnetic mode solver and yield nearly identical effective indices and dispersion curves. The fixed-mode-number approach requires slightly longer computation time but uses half as much memory, making it the preferred choice for memory-limited computers.
}
\section{Introduction}
Dielectric ring resonators are central building blocks of integrated photonic circuits, underpinning applications in optical filtering, modulation, nonlinear frequency conversion, sensing, and quantum photonics \cite{vahala2003optical, bogaerts2012silicon, kippenberg2018dissipative, LN2017, zhang2019broadband, SimsekFDFDsolver}. Their operation depends on the whispering-gallery (WG) modes that circulate around the resonator periphery through total internal reflection. Accurate computation of the resonant frequencies, effective indices, and field profiles of these modes is an essential step in resonator design and in downstream analyses such as dispersion engineering for optical frequency comb generation \cite{kippenberg2018dissipative, SimsekFDFDsolver}.

Three-dimensional (3D) electromagnetic simulation of a ring resonator is computationally expensive: the circumference is typically several tens of wavelengths while the waveguide cross-section is sub- or near-wavelength, so an enormous number of degrees of freedom would be needed to resolve both scales at once. The rotational symmetry of the structure removes this difficulty. Expanding the fields in azimuthal Fourier harmonics separates the 3D problem into a family of independent two-dimensional (2D) problems on the meridional cross-section, one for each azimuthal mode number $m$. This is the basis of the so-called $2.5$-dimensional finite element method (FEM), which has been applied to axisymmetric waveguide and resonator problems in various forms \cite{jin2002finite, Aubourg1991, oxborrow2007traceable}.
Vector discretizations built from N\'{e}d\'{e}lec edge elements \cite{nedelec1980mixed,webb1993edge} avoid the spurious modes that arise from naive nodal discretization of the curl--curl operator \cite{boffi2000,sun2001,monk2003finite,boffi2013}, and a hybrid edge--nodal element, in which the azimuthal field component is represented separately from the transverse ones, was introduced for the cylindrical vector wave equation by Aubourg and Guillon \cite{Aubourg1991} and later applied to WG resonators by Oxborrow \cite{oxborrow2007traceable}. A closely related mixed formulation for dielectric waveguides is given in \cite{koshiba1994vectorial}. The present work does not claim novelty in these foundational ingredients. Its contribution is instead a complete, explicit, and openly released implementation of the method, together with a discussion of a point that is frequently glossed over in the literature: the azimuthal mode number $m$ is, by physical necessity, an integer, yet the practical problem a designer faces is almost always posed in terms of a continuous quantity (an incident wavelength or frequency) that will not in general coincide with any of the discrete resonances the ring supports. We describe two complementary computational strategies for resolving this mismatch and obtaining the effective index at a prescribed wavelength, discuss the reasoning behind each, and compare their respective advantages. A complete, annotated MATLAB implementation is released as open source, together with numerical examples for silicon nitride ring resonators, so that the method is immediately reproducible without recourse to commercial software such as COMSOL Multiphysics. 

\section{Governing Equations}
\begin{figure}[h]
\centering
\includegraphics[width=0.7\textwidth]{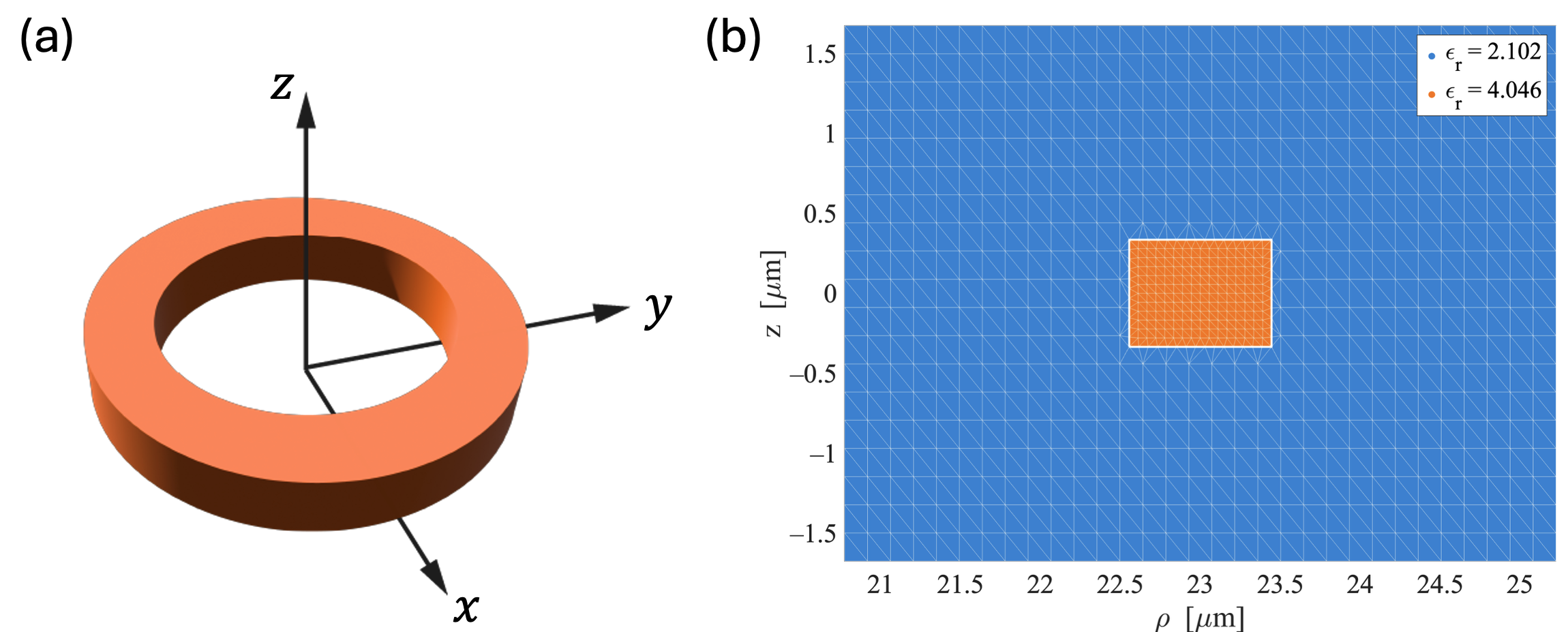}
\captionsetup{width=\textwidth}
\caption{(a) 3D schematic of the dielectric ring resonator. (b) Finite element mesh of the meridional half-plane $(\rho,z)$.}
\label{fig:geometry}
\end{figure}

We consider a dielectric ring resonator that is rotationally symmetric about the $z$-axis, as shown in Fig.~\ref{fig:geometry} (a), consisting of one or more isotropic, lossless dielectric regions of real relative permittivity, $\varepsilon_r \geq 1$.

Under the time-harmonic convention $e^{j\omega t}$, eliminating the magnetic field from the source-free Maxwell equations yields the vector wave equation, 
\begin{equation}
  \nabla\times(\nabla\times\mathbf{E}) - k_0^2\,\varepsilon_r\,\mathbf{E} = \mathbf{0},
  \label{eq:strongform}
\end{equation}
where $k_0^2=\omega^2/c_0^2$ is the eigenvalue of interest and the resonant frequency follows from $f=c_0k_0/(2\pi)$. Under the assumption of homogeneous Dirichlet boundary condition described below, the operator is self-adjoint. Because the structure is axisymmetric, the fields are expanded in azimuthal Fourier harmonics. For azimuthal mode number $m\in\mathbb{Z}$, $\mathbf{E}(\rho,\phi,z) = \widetilde{\mathbf{E}}(\rho,z)\,e^{jm\phi}$ and 
 $ \widetilde{\mathbf{E}}
  = \begin{bmatrix}E_\rho(\rho,z) \ E_\phi(\rho,z) \ E_z(\rho,z)\end{bmatrix}^T,$
so that $\partial/\partial\phi\to jm$. The cylindrical curl then takes the
form
\begin{equation} \nabla_m\times\mathbf{E} =
  \begin{bmatrix}
    \dfrac{jm}{\rho}E_z - \dfrac{\partial E_\phi}{\partial z}\\[0.7em]
    \dfrac{\partial E_\rho}{\partial z} - \dfrac{\partial E_z}{\partial\rho}\\[0.7em]
    \dfrac{\partial E_\phi}{\partial\rho}+\dfrac{E_\phi}{\rho}
    -\dfrac{jm}{\rho}E_\rho
  \end{bmatrix},
  \label{eq:curl}
 \end{equation}
which depends on $\rho$ and $z$ only, reducing the 3D problem to a sequence of independent 2D problems on the meridional half-plane, one per value of $m$. Integrality of $m$ is not a numerical convention but a direct consequence of requiring the field to be single-valued after one full revolution, $\mathbf{E}(\rho,\phi+2\pi,z)=\mathbf{E}(\rho,\phi,z)$. A ring resonator therefore does not support an arbitrary resonant wavelength:
it supports only the discrete set of wavelengths $\lambda_0(m)$, $m\in\mathbb{Z}$, for which a self-consistent, single-valued mode exists. As a rough guide, the azimuthal mode number of the resonance nearest a given wavelength can be estimated from the phase-matching relation
$m\approx n_{\mathrm{eff}}k_0R_c$, where $n_{\mathrm{eff}}$ is an approximate effective index and $R_c$ is the central radius of the ring resonator.

Let $\mathbf{F}=(F_\rho, F_\phi, F_z)$ be a vector test function in the same function space as $\mathbf{E}$. Multiplying \eqref{eq:strongform} by $\mathbf{F}^*$ and integrating over the meridional domain $\Omega$ yields
\begin{equation}
\int_\Omega \mathbf{F}^*\cdot
\bigl[\nabla\times(\nabla\times\mathbf{E})\bigr]\rho\,d\Omega
- k_0^2\int_\Omega\varepsilon_r\,\mathbf{F}^*\cdot\mathbf{E}\,\rho\,d\Omega
= 0.
\end{equation}
Applying the vector curl--curl Green's identity transfers one curl from the trial field to the test field, introducing a boundary term $\int_{\partial\Omega}(\mathbf{n}\times\mathbf{F}^*)\cdot (\nabla_m\times\mathbf{E})\,\rho\,dS$. The computational domain $\Omega$ is a rectangular meridional window chosen large enough that the WG field decays to negligible amplitude before reaching $\partial\Omega$. A homogeneous Dirichlet condition $\mathbf{n}\times\mathbf{E}=\mathbf{0}$ is imposed on all exterior edges, which annihilates the boundary term exactly. The resulting weak formulation is
\begin{equation}
  \underbrace{\int_\Omega
    (\nabla_m\times\mathbf{F})^*\cdot(\nabla_m\times\mathbf{E})
    \,\rho\,d\Omega}_{a(\mathbf{F},\mathbf{E})}
  = k_0^2
  \underbrace{\int_\Omega
    \varepsilon_r\,\mathbf{F}^*\cdot\mathbf{E}
    \,\rho\,d\Omega}_{b(\mathbf{F},\mathbf{E})}.
  \label{eq:weakform}
\end{equation}
The sesquilinear form $a(\mathbf{F},\mathbf{E})$ is linear in $\mathbf{E}$ and conjugate-linear in $\mathbf{F}$, and this conjugation must be applied explicitly when evaluating each matrix block. The form $b(\mathbf{F},\mathbf{E})$ is Hermitian and positive definite.

On each triangular element with vertices $(\rho_1,z_1)$,$(\rho_2,z_2)$,$(\rho_3,z_3)$, area $A$, and barycentric coordinates $L_1,L_2,L_3$, the barycentric-coordinate gradients are
$\nabla L_i=[c_i;b_i]$, with $c_i=\partial L_i/\partial\rho=(\rho_k-\rho_j)/(2A)$ and
$b_i=\partial L_i/\partial z=(z_j-z_k)/(2A)$ for $(i,j,k)$ cyclic. The transverse field components $(E_\rho,E_z)$ are approximated in $H(\mathrm{curl})$ using first-order N\'{e}d\'{e}lec edge elements \cite{nedelec1980mixed,monk2003finite,boffi2013}. For the edge connecting local nodes $i$ and $j$, with global orientation sign $s_{ij}\in\{+1,-1\}$, the edge basis vector is
\begin{equation}
  \mathbf{N}_{ij}
  = s_{ij}(L_i\,\nabla L_j-L_j\,\nabla L_i)
  \equiv\begin{bmatrix}N^\rho_k\\N^z_k\end{bmatrix},
  \label{eq:nedelec}
\end{equation}
and its scalar 2D curl, constant over the element, is $\gamma_k=s_{ij}\cdot2(b_ic_j-b_jc_i)$. The key property of N\'{e}d\'{e}lec elements is that the tangential component of $\mathbf{N}_{ij}$ is continuous across element edges while the normal component may be discontinuous, exactly matching the physical continuity conditions for the tangential electric field at dielectric interfaces \cite{monk2003finite}; this built-in conformity is what eliminates the spurious modes that appear in nodal discretizations of the curl--curl operator \cite{boffi2000,sun2001}. The azimuthal component $E_\phi$ is approximated in $H^1$ by standard
piecewise-linear nodal (Lagrange) basis functions $L_a$, since it does not require the same tangential-continuity treatment as the transverse components. The element degree-of-freedom vector is $\mathbf{x}_e=[u_1,u_2,u_3,p_1,p_2,p_3]^T$, where $u_k$ are edge circulations
of the transverse field and $p_a$ are nodal values of $E_\phi$.

Testing and expanding \eqref{eq:weakform} with these bases, and applying the complex conjugation required by the sesquilinear form to the test-function curl, produces four stiffness blocks. The edge--edge, edge--nodal coupling, and nodal--nodal blocks are
\begin{align}
  K^{ee}_{ij}
  = & \int_e\!\left[\gamma_i\gamma_j
    +\frac{m^2}{\rho^2}\bigl(N^z_iN^z_j+N^\rho_iN^\rho_j\bigr)
    \right]\rho\,d\Omega,
  \label{eq:Kee} \\
  K^{ep}_{ij}
  = & \int_e jm\!\left[b_jN^z_i
    +\!\left(c_j+\frac{L_j}{\rho}\right)\!N^\rho_i\right]\!d\Omega,
  \label{eq:Kep} \\
  K^{pp}_{ij}
  = & \int_e\!\left[b_ib_j
    +\!\left(c_i+\frac{L_i}{\rho}\right)
     \!\!\left(c_j+\frac{L_j}{\rho}\right)
    \right]\!\rho\,d\Omega,
  \label{eq:Kpp}  
\end{align}
where $K^{pe}_{ij}=\overline{K^{ep}_{ji}}$. Because the edge and nodal bases represent different field components and are mutually orthogonal in the vector inner product, the element mass matrix is block diagonal,
$M_e=\bigl[\begin{smallmatrix}M^{ee}&0\\0&M^{pp}\end{smallmatrix}\bigr]$, with
\begin{align}
  M^{ee}_{ij}
  &= \int_e\varepsilon_r\,(N^\rho_iN^\rho_j+N^z_iN^z_j)\,\rho\,d\Omega,
  \label{eq:Mee}\\
  M^{pp}_{ij}
  &= \int_e\varepsilon_r\,L_iL_j\,\rho\,d\Omega,
  \label{eq:Mpp}
\end{align}
both real, symmetric, and positive definite. All element integrals are evaluated by a seven-point Gaussian quadrature rule. Elements are assembled into global matrices $K=\bigcup_eK_e$ and $M=\bigcup_eM_e$ by the standard finite-element connectivity scatter, yielding the global generalized eigenproblem
\begin{equation}
  K(m)\,\mathbf{x} = k_0^2\,M\,\mathbf{x}, 
  \label{eq:gep}
\end{equation}
in which $K(m)$ depends on the azimuthal mode number both quadratically, through \eqref{eq:Kee}, and linearly, through \eqref{eq:Kep}, while the mass matrix $M$ does not depend on $m$ at all. \eqref{eq:gep} can be solved efficiently with a shift-invert Krylov method (\texttt{eigs} in MATLAB), with the shift chosen near the operating wavelength so that only a small number of eigenvalues closest to the shift need be resolved from a single sparse factorization. This can be done using two approaches.

The first approach leaves the governing equation exactly as derived: for multiple trial integer values of $m$ bracketing the phase-matching estimate, the linear generalized eigenproblem in \eqref{eq:gep} is solved directly, yielding a discrete set of resonant wavelengths $\lambda_0(m)$ and their associated effective indices $n_{\mathrm{eff}}(m)=mc_0/(\omega R)$. Because $m$ only ever appears as a fixed parameter of the matrix $K(m)$, this recovers, for each trial $m$, a well-conditioned Hermitian eigenproblem of exactly the same size and structure derived above.
The trial values of $m$ used to bracket the target wavelength should be selected adaptively rather than fixed in advance. An initial value is obtained from the phase-matching estimate, $m=\mathrm{round}(n_{\mathrm{est}}k_0R_c)$; \eqref{eq:gep} is then solved at this $m$ for the $i$th mode of interest, yielding a resonant wavelength $\lambda_{\mathrm{res}}^{(i)}$. If $\lambda_{\mathrm{res}}^{(i)}$ exceeds the target wavelength $\lambda_{\mathrm{target}}$, progressively smaller integer
values of $m$ are tried until the corresponding resonant wavelength $\lambda_{\mathrm{res}}^{(i)*}$ falls below $\lambda_{\mathrm{target}}$; conversely, if $\lambda_{\mathrm{res}}^{(i)}$ falls below $\lambda_{\mathrm{target}}$, progressively larger values of $m$ are tried until $\lambda_{\mathrm{res}}^{(i)*}$ exceeds it. This procedure follows directly from the phase-matching relation, since increasing $m$ shortens the resonant wavelength and decreasing $m$ lengthens it. Once two integer values of $m$ have been found whose resonant wavelengths bracket $\lambda_{\mathrm{target}}$ from above and below, the corresponding pair of effective indices, $(\lambda_{\mathrm{res}}^{(i)},n_{\mathrm{eff}}^{(i)})$
and $(\lambda_{\mathrm{res}}^{(i)*},n_{\mathrm{eff}}^{(i)*})$, are interpolated to obtain the effective index at $\lambda_{\mathrm{target}}$. 
 
The second approach inverts the roles of the two variables: the incident wavelength is fixed from the outset, so that $\lambda=k_0^2$ is a known constant, and the azimuthal mode number $m$ is instead treated as the
unknown to be solved for. Because $K(m)$ depends on $m$ through both a linear and a quadratic term, fixing $\lambda$ converts \eqref{eq:gep} into a quadratic eigenvalue problem in $m$,
\begin{equation}
  \bigl[m^2A_2 + mA_1 + A_0(\lambda)\bigr]\,\mathbf{x} = \mathbf{0},
  \label{eq:qep}
\end{equation}
which is solved by linearizing it \cite{SimsekDieWGsolver} into a generalized eigenproblem of twice the original size and applying the same shift-invert eigensolver, seeded near the phase-matching estimate of $m$. The eigenvalue returned by this solve is, in general, a real but non-integer number, since nothing in the underlying linear algebra enforces integrality. It is simply the value of $m$ for which a self-consistent field solution would exist at exactly the prescribed wavelength, were $m$ permitted to vary continuously. 

\section{Numerical Results}
\begin{figure}[h]
    \centering
    \includegraphics[width=0.9\linewidth]{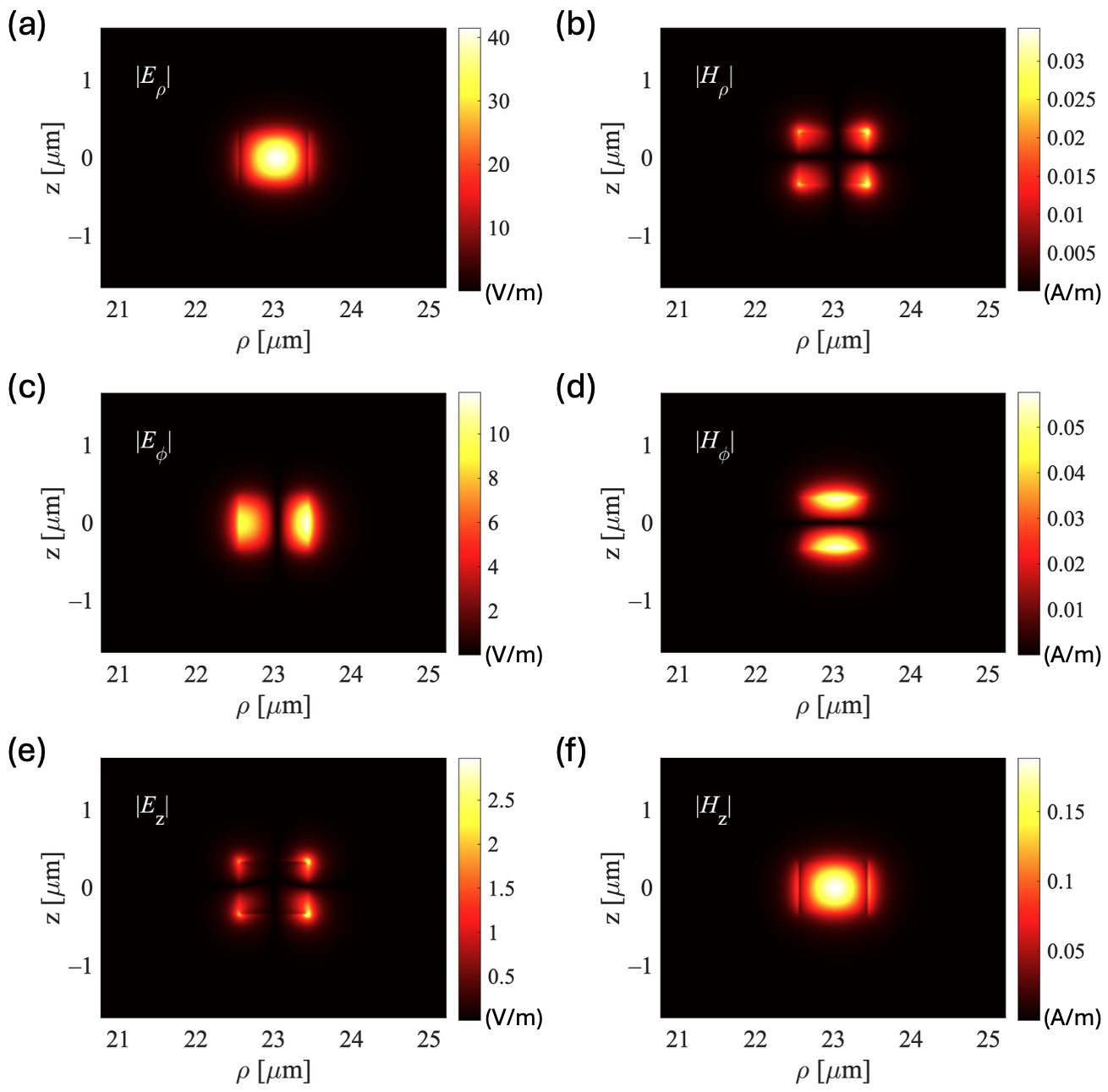}
    \caption{Electric and magnetic field distributions for the first resonant mode. The central radius, width, and height of the ring resonator are 23 $\mu$m, 890 nm, and 670 nm, respectively. The excitation wavelength is 1.06 $\mu$m. }
    \label{figure_fields}
\end{figure}

To compare the accuracy and efficiency of both approaches, we study a Si$_3$N$_4$ \cite{Si3N4Sellmeier} ring resonator surrounded by SiO$_2$ \cite{Palik}. The central radius, width, and height of the resonator are 23 $\mu$m, 890 nm, and 670 nm, respectively. For the first study, the excitation wavelength is 1.06 $\mu$m. Figure \ref{figure_fields} shows the electric and magnetic field distributions for the first resonant mode. We observe that the electric field density reaches a maximum at $\rho = 23.44~\mu$m, slightly off-centered, as expected.  Using either approach with a high-quality mesh with 80 elements per wavelength (EPW), the effective index of the ring is calculated to be 1.85804, which is very close to the value (1.85806) determined with COMSOL Multiphysics.  

\begin{figure}[h]
\centering
\includegraphics[width=0.7\textwidth]{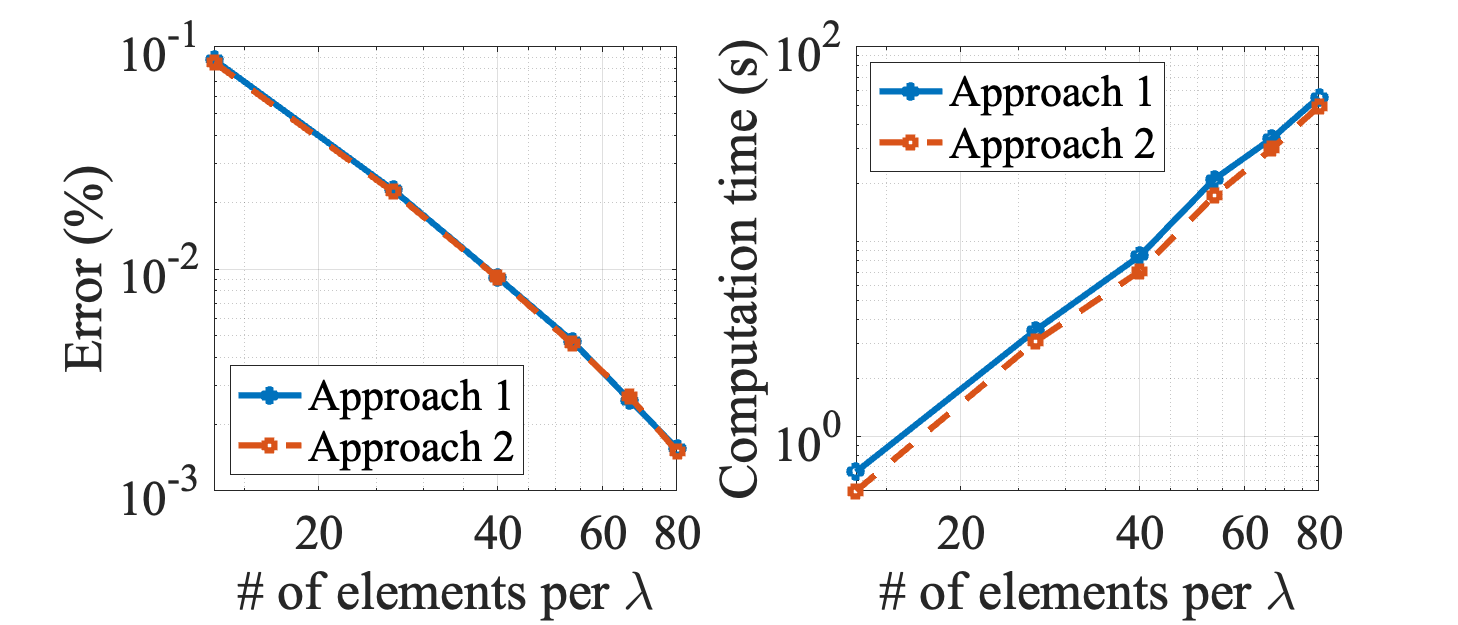}
\caption{Convergence of the effective index and computation time of approaches 1 and 2 with mesh density, expressed as the number of elements per wavelength (EPW). Left: relative error in effective index compared with a COMSOL Multiphysics reference value, decreasing with a log-log slope of $-2$. Right: total computation time, increasing with a log-log slope of $2.5$.}
\label{fig:convergence}
\end{figure}

To quantify convergence, both approaches were run at $\lambda_0=1.06\ \mu\text{m}$ for the same ring resonator at six mesh densities spanning approximately 10--80 elements per wavelength (EPW). Figure~\ref{fig:convergence} shows the
relative error in effective index and total computation time versus number of EPW on logarithmic axes. The error decreases with a slope of $\approx-2$ for both approaches, close to the $O(h^2)$ superconvergence expected for eigenvalues from a variational edge-element formulation. Computation time grows with a slope of $\approx2.5$ for both
approaches, between the $\mathrm{EPW}^2$ scaling expected if element-matrix assembly dominates and the $\mathrm{EPW}^3$ scaling expected if sparse factorization dominates, indicating both contribute over this mesh range.
 
These results show that ring-resonator modal analysis can be carried out equally well by fixing $m$ and interpolating, or by fixing the wavelength and solving for $m$ directly. The two approaches agree with each other and with COMSOL Multiphysics with an error less than 0.01 \% when 40 or a higher EPW is used. Their overall computation times are also comparable, but the first approach takes slightly longer computation time than the second approach. In theory, a single fixed-wavelength solution should take three to four times longer than the first approach, since the fixed-wavelength matrices have twice the linear dimension of the fixed-$m$ matrices, and since both assembly and LU factorization scale faster than linearly with matrix size. However, this second approach needs only one solve, while the fixed-$m$ approach needs at least two, and sometimes several, to bracket the target wavelength before interpolating. This is why total runtimes end up similar. Of course, parallelization or vectorization could speed up either approach. Memory is where the two approaches differ sharply. At the finest mesh considered (80 EPW), peak memory usage was about 400~MB for the fixed-$m$ approach and 900~MB for the fixed-wavelength approach, roughly a factor of two, due to the larger fill-in from LU factorization of the doubled-size matrix pencil. We therefore recommend the fixed-$m$ approach on memory-limited computers, and the fixed-wavelength approach when memory is not a concern and only a single target wavelength is needed.

\begin{figure}[h]
    \centering
    \includegraphics[width=0.8\linewidth]{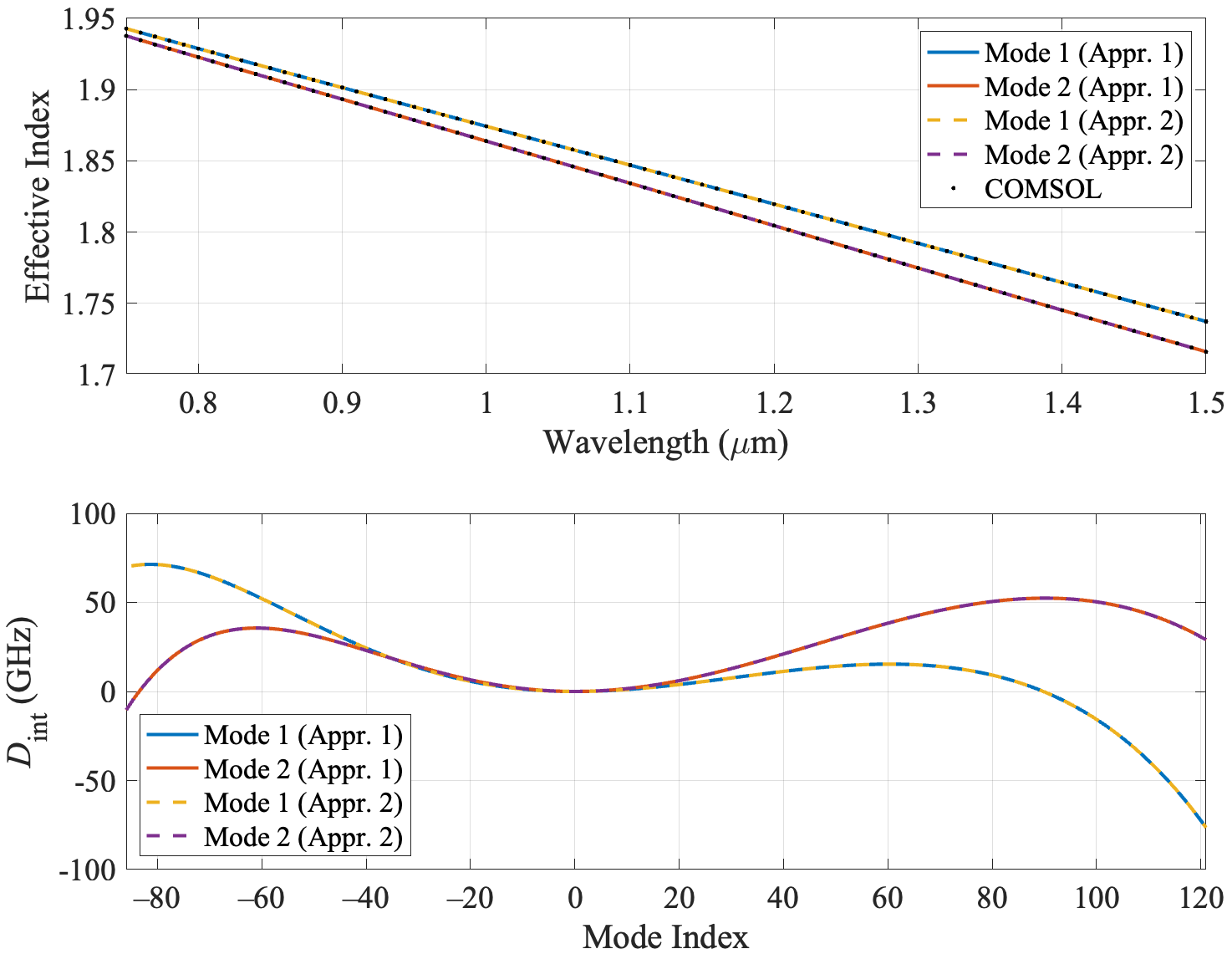}
    \caption{(a) Effective indices vs. excitation wavelength and (b) integrated dispersions of the first two modes, assuming a pump wavelength of 1060 nm for a Si$_3$N$_4$ ring resonator surrounded by SiO$_2$. Solid and dashed curves are the outputs of approaches 1 and 2, respectively.}
    \label{fig2}
\end{figure}

As an example application of the developed solvers, we use the same ring resonator and calculate the effective indices of the first two resonant modes across 76 wavelengths spanning $0.75$--$1.50\ \mu\text{m}$. The solid and dashed curves in Fig. \ref{fig2} (a) show the indices computed by approaches 1 and 2, respectively. Both approaches were run on identical meshes with 80 EPW. In the same figure, the dotted curves show the effective indices of the same modes obtained with COMSOL Multiphysics. We observe very good agreement among the solutions.
Figure \ref{fig2} (b) shows the integrated dispersions assuming a pump wavelength of 1.06 $\mu$m calculated with the recipe provided in \cite{Simsek1stopshop}. We observe anomalous dispersion for the second mode over the entire wavelength range.
 
\section{Conclusion}
We have presented an axisymmetric, hybrid edge--nodal finite element solver for the whispering-gallery modes of dielectric ring resonators, together with two complementary strategies for reconciling the discrete, integer-valued azimuthal mode number with the continuous incident wavelength at which a device is actually operated. The first approach solves the
standard linear generalized eigenproblem at fixed integer $m$ and interpolates the resulting effective index to the target wavelength; the second fixes the wavelength and solves, via linearization, a quadratic eigenproblem for $m$ directly. Both approaches were validated against COMSOL Multiphysics and shown to agree with it, and with each other, to within a few parts in $10^4$ in effective index when a high-quality mesh (e.g., 80 elements per wavelength) is used. The effective-index convergence rates were found to follow the expected $O(h^2)$ superconvergence behavior of a variational edge-element formulation. The two approaches were found to require comparable overall computation time, since the higher per-solve cost of the fixed-wavelength approach is offset by the additional bracketing solves required by the fixed-$m$ approach. Their memory requirements, however, differ significantly, with the fixed-wavelength approach consuming nearly two times the peak memory of the fixed-$m$ approach at the mesh densities considered. The fixed-$m$ approach is accordingly recommended for computers with limited memory, while the fixed-wavelength approach remains preferable when a single targeted wavelength is of interest and memory is not a constraint. Complete, open-source MATLAB implementations of both approaches, together with the examples presented here, are made available so that the method and this comparison are directly reproducible.

\subsection*{Data Availability Statement} 
All the data presented in this work and all the codes to produce these results can be found at \url{https://github.com/simsekergun/fem_ring_modesolver}.

\clearpage
\bibliographystyle{unsrt}
\bibliography{refs}
\end{document}